\pdfoutput=1
\documentclass[11pt,a4paper]{article}
\usepackage{array}
\usepackage[footnotesize,bf]{caption}
\usepackage[dvips]{graphicx}
\usepackage{color} 
\usepackage{amsmath}
\usepackage{fancyhdr}
\usepackage[]{natbib}
\usepackage[LGRgreek]{mathastext}
\usepackage[yyyymmdd,hhmmss]{datetime}

\fancyheadoffset[R]{0pt}
\fancyfootoffset[R]{0pt}

\definecolor{orange}{rgb}{1.0,0.5,0.0}
\definecolor{aqgr}  {rgb}{0.0,1.0,0.6} 
\definecolor{viol}  {rgb}{0.8,0.6,0.8}
\definecolor{figdr} {rgb}{1.0,1.0,1.0} 
\definecolor{colnu} {rgb}{1.0,0.0,1.0} 
\definecolor{colhd} {rgb}{1.0,0.8,0.0} 

\newcolumntype{C}[1]{>{\centering\let\newline\\\arraybackslash\hspace{0pt}}m{#1}}
\newif\ifhpar

\title{\vspace{-0.5cm}\bfseries{\textsc{A deep learning-inspired model 
   \\ of the hippocampus as storage device \\ of the brain extended dataset}}}
   
\author{Alessandro Fontana} \date{}

\begin{document}
\maketitle
   
\clubpenalty=10000
\widowpenalty=10000

\begin{abstract}
The standard model of memory consolidation foresees that memories are initially recorded in the hippocampus, while features that capture higher-level generalisations of data are created in the cortex, where they are stored for a possibly indefinite period of time. Computer scientists have sought inspiration from nature to build machines that exhibit some of the remarkable properties present in biological systems. One of the results of this effort is represented by artificial neural networks, a class of algorithms that represent the state of the art in many artificial intelligence applications. In this work, we reverse the inspiration flow and use the experience obtained from neural networks to gain insight into the design of brain architecture and the functioning of memory. Our starting observation is that neural networks learn from data and need to be exposed to each data record many times during learning: this requires the storage of the entire dataset in computer memory. Our thesis is that the same holds true for the brain and the main role of the hippocampus is to store the ``brain dataset'', from which high-level features are learned encoded in cortical neurons.    
\end{abstract}


\section{Memory and learning systems} 

\ifhpar \colorbox{colhd}{different types of memory and learning} \\ \fi
In the brain different types of memory are implemented. Based on duration, memory can be classified in short-term (or working) memory and long-term memory \citep{baddely2007}. Based on content, memory is classified in declarative (further subdivided in episodic and semantic) and implicit or procedural \citep{squire2009}. Regarding learning, the number of ways in which the brain can learn exceeds the power of any classification system.  

\ifhpar \colorbox{colhd}{hippocampus and cortex} \\ \fi
The hippocampus (a seahorse-shaped brain structure located in the medial temporal lobe) and the cortex (a 3 mm-thick layer of tissue distributed on the surface on the brain)  are both involved in the process of memory formation. Overall, the empirical evidence seems to hint that the hippocampus stores complete and unprocessed memory records for a short time, while the cortex develops features capturing high-level generalisations of data \citep{preston2013}, that are stored for longer (possibly indefinite) periods. These generalisations may correspond to the schemas of Piaget's developmental theory \citep{piaget1952}. 

\begin{figure}[t]
\begin{minipage}[t]{0.47\textwidth} \centering \hspace*{-0.0cm}
{\fboxrule=0.0mm\fboxsep=0mm\fbox{\includegraphics[width=0.90\textwidth]{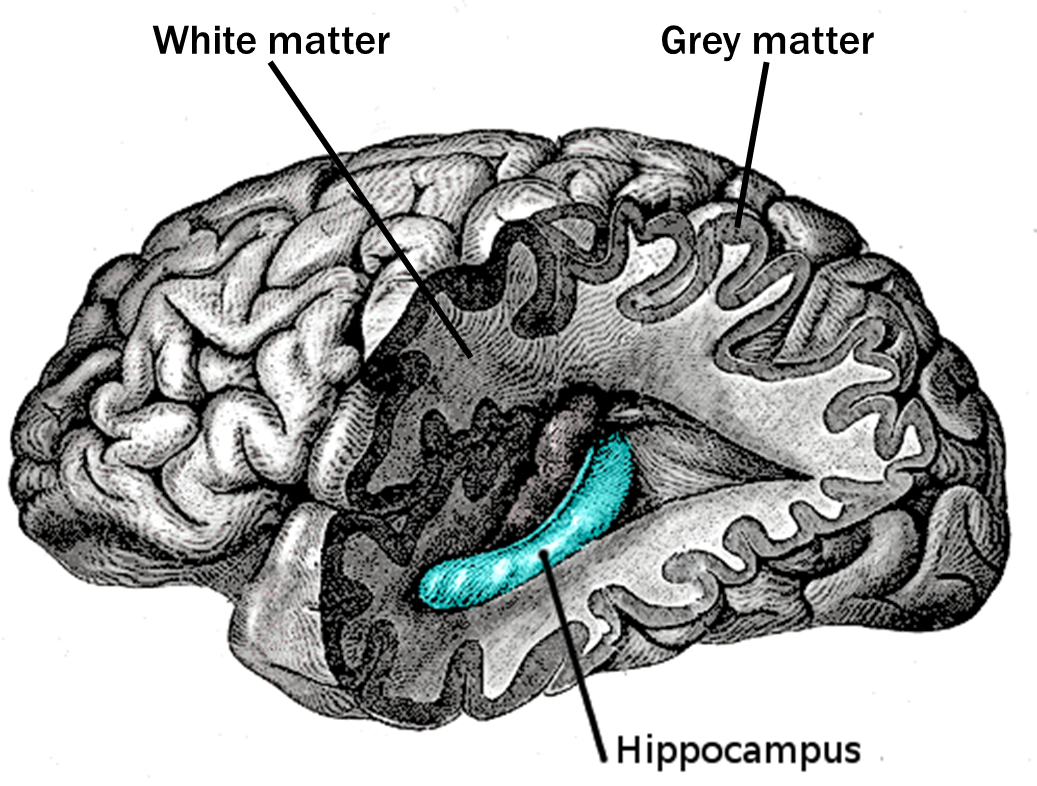}}}
\caption{Drawing of the human brain showing disposition of grey matter (on the cortex and in the depth of the brain) and white matter (in the intermediate space), as well as location of the hippocampus.}
\label{brain}
\end{minipage} \qquad
\begin{minipage}[t]{0.47\textwidth} \centering \hspace*{-0.0cm}
\includegraphics[width=1.00\textwidth]{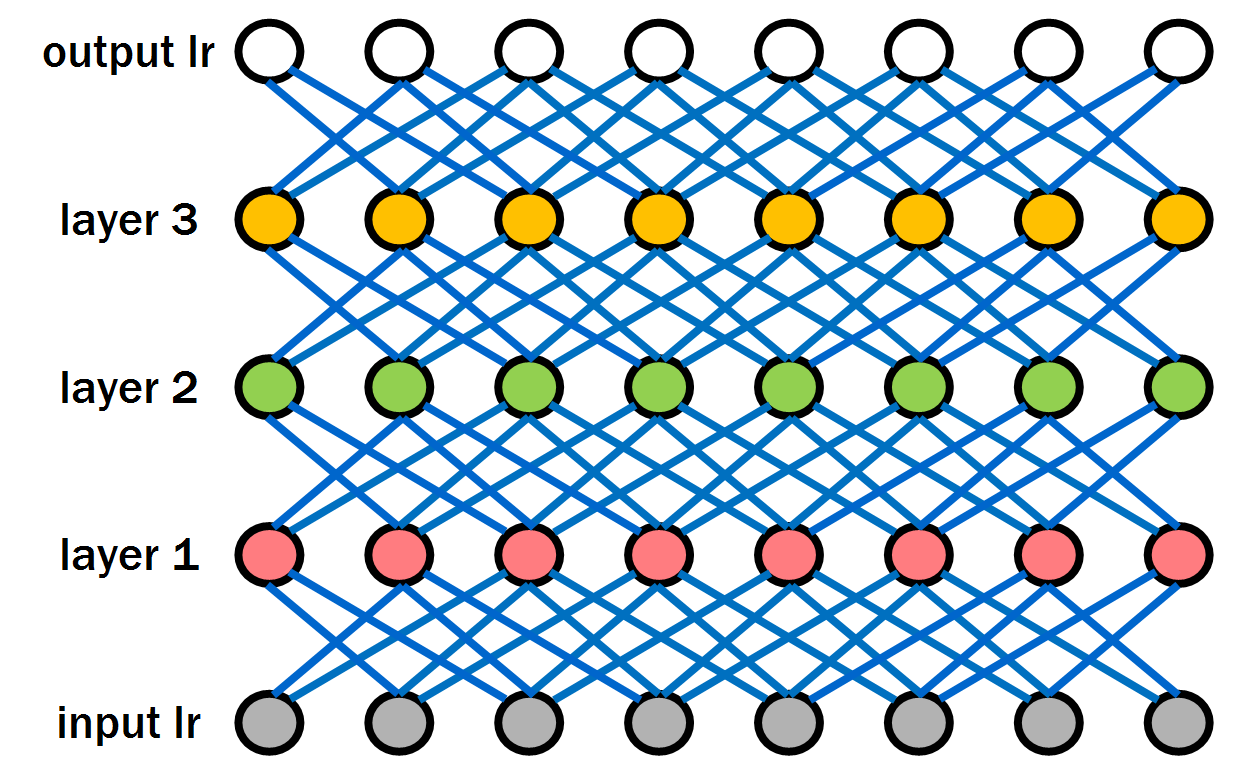}
\caption{Structure of a deep neural network. The network is composed of computational units called neurons, arranged in layers stacked on top of each other (input and output layers are always present, the number of intermediate layers defines the network's depth). This structure can also be regarded as an alternation of grey and white layers.}
\label{deepnet}
\end{minipage} \qquad
\end{figure}

\ifhpar \colorbox{colhd}{neural networks} \\ \fi
Artificial neural networks (simply neural networks from now on) are computational models that take inspiration from the structure of biological neural networks. A deep neural network (Fig.~\ref{deepnet}) is a particular type of neural network (called ``multilayer perceptron'') composed of a number of computational units called neurons, grouped in layers stacked on top of each other (input and output layers are always present, the number of intermediate layers defines the network's depth). 

\ifhpar \colorbox{colhd}{deep learning, convolutional} \\ \fi
Deep neural networks represent the state of the art in many artificial intelligence applications, such as computer vision, speech recognition, natural language processing. For visual classification tasks, the best results are achieved by convolutional neural networks trained with ``back-propagation'' algorithm \citep{krizhevsky2012}, which appears to be less prone to the vanishing gradient problem \citep{hochreiter2004} when huge amounts of data are used. A recent method \citep{kaiming2015} has achieved human-like performance on the ImageNet dataset using a network with 152 layers.

\ifhpar \colorbox{colhd}{mismatch} \\ \fi
Neural networks work well in a growing number of applications, yet their structure seems to lack biological plausibility. If neural networks were a good model for brain architecture, the brain should be characterised by a striped colour pattern, with an alternation of grey and white regions. The topological structure of Fig.~\ref{deepnet} could also map to an irregular, spaghetti-like geometrical structure in the brain: in this case, we would expect to observe a uniform (noisy) colour pattern. What we see, instead, is a distribution of grey matter concentrated on the brain's surface and in the central core, with the intermediate regions filled by white matter (Fig.~\ref{brain}).

\ifhpar \colorbox{colhd}{objective and structure} \\ \fi
Despite these apparent contradictions, we will argue that the structure of neural networks and the functioning of the associated algorithms are fully compatible with brain architecture. The rest of the paper is organised as follows: based on the known facts about the working of biological memory summarised in section 2 and on the functioning of neural networks described in section 3, we will present a proposal for brain architecture in section 4, which includes an unsuspected role for the hippocampus; section 5 discusses some implications for memory and learning in normal and pathological conditions; section 6 draws the conclusions and outlines future research directions. This work is intended for a readership of both neuro- and computer scientists: therefore concepts are presented with sufficient background information.

\section{Memory and learning in the brain: hippocampus and cortex}  

\ifhpar \colorbox{colhd}{hippocampus} \\ \fi
Hippocampi are located (more or less) in the centre of each brain hemisphere, in a region where fibres carrying multiple sensory inputs converge \citep{amaral1995}. Each hippocampus is directly connected to a portion of cortex called entorhinal cortex, which functions as a connection hub to and from the rest of cortex. The size of the entorhinal cortex is relatively small and (presumably) only allows a subset of cortical fibres to reach the hippocampus \citep{canto2008}.  

\ifhpar \colorbox{colhd}{patient H.M.} \\ \fi
A decisive contribution to our understanding of memory processes came from studies conducted on patient H.M., who had large portions of both medial temporal lobes (including the hippocampi) removed at age 27, in an attempt to cure severe epilepsy. The surgical procedure was successful in treating epilepsy, but left he patient with an almost complete anterograde amnesia and a graded retrograde amnesia. In other words, H.M. was unable to form new declarative memories and his recollection of recent past events was impaired, while older memories were intact \citep{scoville1957}.   
 
\ifhpar \colorbox{colhd}{standard model} \\ \fi
Based on these and other studies, the ``standard model'' of memory formation and consolidation was proposed and consolidated \citep{squire1986, mcclelland1995}. The model foresees that new (declarative) memories are first stored in the hippocampus and then, in a process that can last decades, gradually transferred to the cortex, where they are stored indefinitely. Once the transfer is complete, memories are retained even if the hippocampus is removed or damaged. 

\ifhpar \colorbox{colhd}{neurogenesis, physical activity} \\ \fi
The vast majority of brain neurons are generated during embryonic development. However, new neurons are created also in specific regions of the adult brain. The generation of neurons in the dentate gyrus of the hippocampus is a process that continues for the whole duration of life \citep{eriksson1998}, favoured by physical exercise \citep{nokia2016}. It is natural to think that such neurons are involved in the process of memory formation in this brain region.  

\ifhpar \colorbox{colhd}{recent model} \\ \fi
A recent model \citep{kitamura2017} suggests that the instantiation of the cortical representation of memory ``engrams'' \citep{bruce2001} occurs from the very beginning. New memories, instead of being first recorded in the hippocampus and then gradually copied or moved to the cortex, would be written in both places in parallel. The cortical representation would be immature at first and develop with time to more mature forms.

\section{Memory and learning in neural networks}  

\begin{figure}[t]
\begin{minipage}[t]{0.47\textwidth} \centering \hspace*{-0.0cm}
\includegraphics[width=1.00\textwidth]{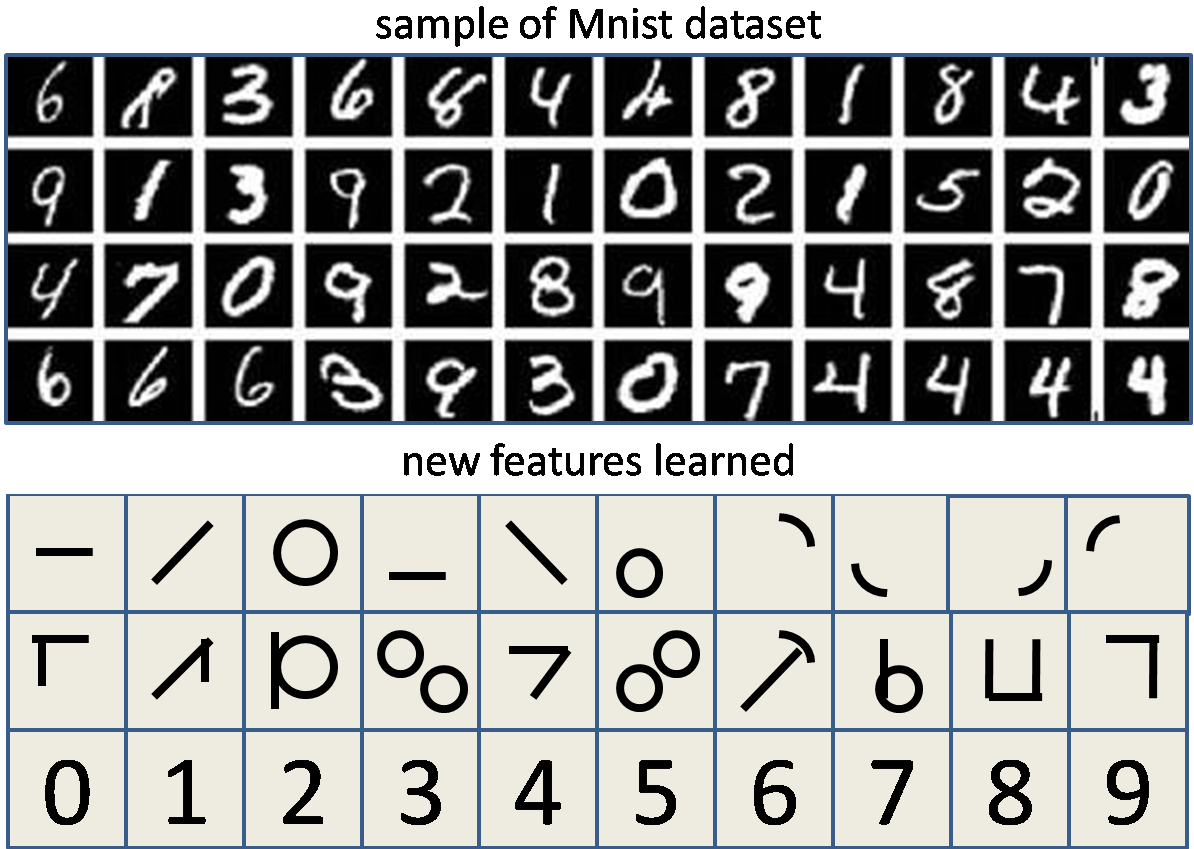}
\caption{Mnist dataset. The upper part of the figure shows a sample of Mnist dataset, consisting of 60000 images of handwritten characters belonging to 10 classes (0-9). The lower part contains stylised examples of new features learned by the network, with increasing abstraction level.}
\label{mnist}
\end{minipage} \qquad
\begin{minipage}[t]{0.47\textwidth} \centering \hspace*{-0.4cm}
\includegraphics[width=1.10\textwidth]{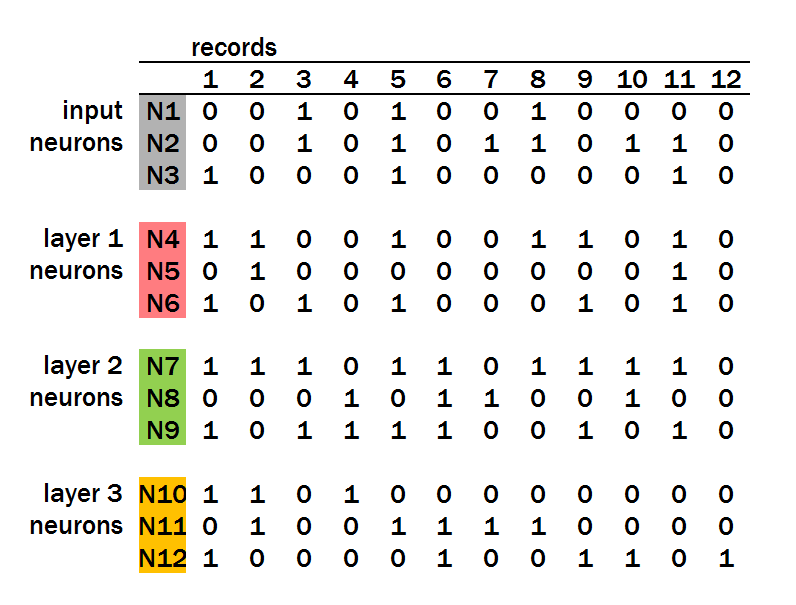}
\caption{Example of extended dataset. Rows correspond to features /neurons, columns correspond to records. Input neurons encode the proper dataset features, neurons of layer 1 and subsequent layers encode new features acquired by the neural network as a result of the learning process.}
\label{dataset}
\end{minipage} \qquad
\end{figure}

\ifhpar \colorbox{colhd}{Mnist dataset} \\ \fi
Neural networks learn from data, organised in a dataset, structured as a collection of records. The upper part of Fig.~\ref{mnist} shows a sample of Mnist \citep{lecun1998}, a dataset commonly used to train neural networks, composed of 60000 images representing hand-written characters from 0 to 9, where each record of the dataset is a set of numbers (the dataset variables) encoding the grey shades of the 28x28 individual pixels that compose an image. From the dataset, the network learns higher-level features, representing oriented edges or simple shapes, combination of simple shapes, and the ``concepts'' of digits (lower part of Fig.~\ref{mnist}).        

\ifhpar \colorbox{colhd}{word dataset} \\ \fi
Datasets can contain any kind of elements: real numbers, categorical values, words, etc. An example of word dataset is shown in Table~\ref{words}, in which records correspond to fruits on sale at the local market in some days. In this case, high-level features may correspond to ``rules'' that occur more frequently than expected by chance, e.g.: ``round fruits occur on odd days'', ``non-round fruits occur on even days'', or ``there is an alternation of occurrence of round fruits and non-round fruits'', etc. 

\begin{table}[t]
\vskip 0.25cm
\centering
\begin{tabular}{c c c c c c c c c}
record nr. & (1)    & (2)       & (3)   & (4)    & (5)   & (6)  & (7)   & (8)     \\
1st word   & orange & pineapple & kiwi  & banana & apple & pear & lemon & avocado \\
2nd word   & march  & march     & april & april  & may   & june & june  & june    \\
3rd word   & 21st   & 18th      & 14th  & 28th   & 5th   & 8th  & 9th   & 30th    \\
\end{tabular}
\vskip 0.25cm
\caption{Example of word dataset, in which records correspond to fruits on sale at the local market in some days (each column corresponds to a different record).}
\label{words}
\end{table}

\ifhpar \colorbox{colhd}{learning} \\ \fi
For learning, dataset variables are mapped to neurons of the neural network input layer, while high-level features are encoded in neurons of successive layers. Learning consists in modifying the parameters that define the strength of connections between neurons which, collectively, represent the memory of the network. Neural networks can learn to perform a variety of tasks (e.g., recognition, classification, etc.): once learning is finished, such tasks can be performed on new data very quickly. 

\ifhpar \colorbox{colhd}{learning procedure} \\ \fi
The typical learning procedure is divided in two parts: training (in which connection parameters are optimised based on the dataset) and test (in which the network is tested on data not used for training). The training part is in turn structured as a cycle composed of two phases: in the ``change'' phase the algorithm brings small changes to connection parameters, while in the ``assessment'' phase the network performance is measured on the dataset, yielding a performance score. This two-phase cycle (called ``epoch'' in the machine learning jargon) is repeated a number of times, until the score reaches a satisfactory value. 

\begin{verbatim}
      Training:
      for i=1 to maxval
         change phase
         assessment phase
      next i
\end{verbatim}

\ifhpar \colorbox{colhd}{supervised /unsupervised} \\ \fi
With ``back-propagation'' (the most commonly used supervised learning algorithm), the score depends on the percentage of records classified correctly for output neurons and on the ``propagated'' gradient of the classification error for neurons of intermediate layers; with unsupervised learning algorithms, other scores are used. While in supervised learning all neurons are optimised at the same time, in unsupervised learning, e.g. ``contrastive divergence'' \citep{hinton2002}, neurons are usually optimised one layer at a time: first layer 1 neurons, then layer 2 neurons, etc. 

\ifhpar \colorbox{colhd}{learning cycle, data structures} \\ \fi
Let us suppose that unsupervised learning is used and layer 1 neurons are being optimised: Fig.~\ref{dataset} shows all data structures needed in the training process. The assessment phase procedure requires that: i) input neurons are exposed to all data records, ii) outputs of layer 1 neurons are calculated for all data records and iii) the score is calculated for the neurons under optimisation. 

\ifhpar \colorbox{colhd}{need to read /write} \\ \fi
This requires that the entire dataset is \textit{read} at each assessment phase during training (first row block in Fig.~\ref{dataset}). Furthermore, computing the outputs of layer 1 neurons (needed to obtain the score) requires that these outputs are also \textit{written} (the second row block in the figure). These additional rows are strictly-speaking not part of the dataset, but correspond to the new features learned, whose values must be calculated for each record (each column). 

\ifhpar \colorbox{colhd}{extended dataset} \\ \fi
It is convenient to define the \textbf{extended dataset} as the union, for each record, of all values of dataset variables and successive layers' neurons. With this definition, we can conclude that the training procedure requires that the extended dataset be accessed \textit{in both reading and writing} at each assessment phase during training. This, in turn, requires the storage of the extended dataset for the entire duration of the process. 

\ifhpar \colorbox{colhd}{need for dataset storage} \\ \fi
Some training algorithms may be able to process data ``on the fly'', with each record seen once and then discarded. In principle this is possible, but the most commonly used algorithms, proved effective in practical applications, need to be exposed to data many times. We will argue that this need is shared also by the ``algorithms'' operating in the brain.  


\section{Implementation of deep networks in the brain}  

\begin{figure}[t]
\begin{minipage}[t]{0.47\textwidth} \centering \hspace*{-0.0cm}
\includegraphics[width=1.00\textwidth]{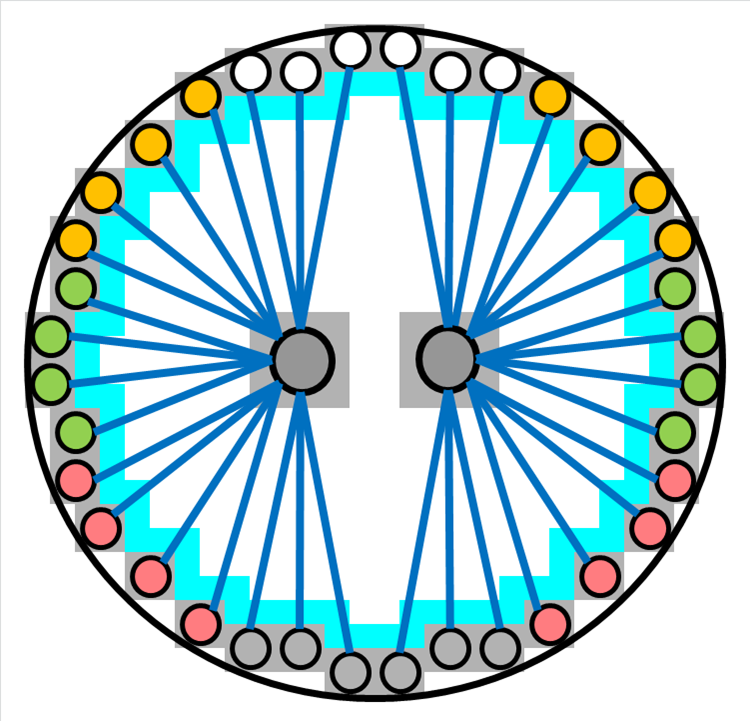}
\caption{Proposed brain architecture. Features are implemented as neurons in the cortex. Neurons are organised in layers, shown with different colours: input (sensory) neurons in grey, layer 1 neurons in red, layer 2 neurons in green, layer 3 neurons in yellow, etc. Layer n neurons' inputs are connected with the outputs of neurons of all previous layers through lateral connections (shown as a light blue belt around the cortex). Neurons send also their axons to the hippocampi (blue radial links).}
\label{ball}
\end{minipage} \qquad
\begin{minipage}[t]{0.47\textwidth} \centering \hspace*{-0.6cm}
\includegraphics[width=1.10\textwidth]{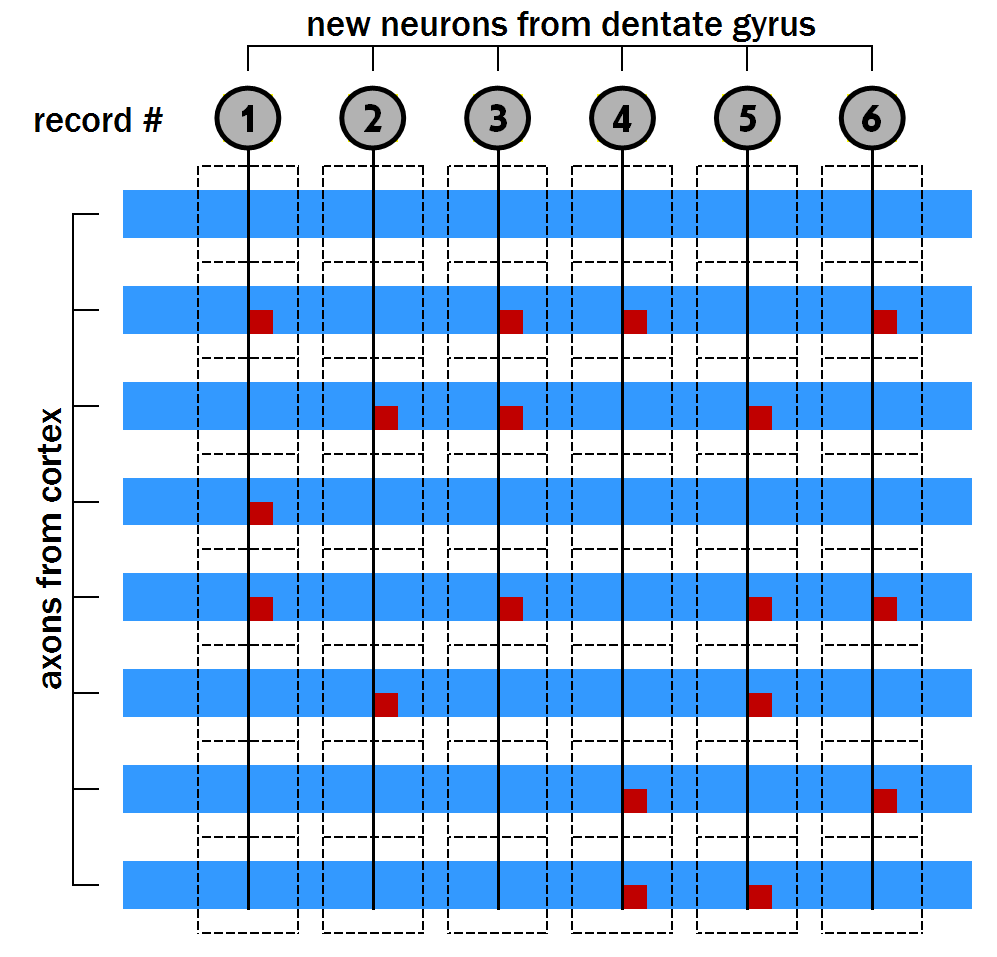}
\caption{Creation of new associations among cortical neurons in the hippocampus. The axons of all cortical neurons, representing as many features, are projected to the hippocampus, where they are very close to each other (much closer than their respective cellular bodies). When a set of neurons fire together, their axons make connections to the dendrites of a new neuron generated in the dentate gyrus: in this way, a new record of the ``brain dataset'' is created.}
\label{hippoc}
\end{minipage} \qquad
\end{figure}

\ifhpar \colorbox{colhd}{features and associations} \\ \fi
Understanding and interpreting reality is the main objective of brain activity, and reality can be conceptualised in terms of \textbf{features}, defined as descriptors or qualities that can be used to characterise a given situation. Features can be simple visual qualities, such as ``red colour'', ``round shape'', ``vertical orientation'', or correspond to more complex visual-motor characteristics, such as ``dance movements'', ``hiding an object with the hand'', etc. Therefore, in our interpretation, the concept of feature encompasses all components of mental life, ranging from simple perceptual elements to the most complex and abstract ideas. The model of reality is constructed by establishing meaningful associations among features: such associations become features in their own right, and can take part in further associations.

\ifhpar \colorbox{colhd}{features and associations, physical implementation} \\ \fi
Our hypothesis is that features of any complexity are implemented in the brain by single neurons, an idea already entertained \citep{quiroga2005}. A neuron may represent a basic visual or auditory feature or the most complex philosophical thought: the complexity of a feature is given by its relation with all other features. We further conjecture that an association among features A, B and C is physically built by recruiting a new neuron N and linking through synapses the axons of neurons A, B and C to the dendrites of N, which encodes a new feature representing the association.    

\ifhpar \colorbox{colhd}{cortical neurons, lateral connections} \\ \fi
Neurons encoding features are located in the cortex (Fig.~\ref{ball}). Visual features, for instance, are mapped to the occipital cortex, auditory features are mapped to the auditory cortex, tactile features are mapped to the parietal cortex, etc. Neurons are connected with each other through ``lateral'' connections (light blue belt around the cortex in the figure) and are arranged in layers. The neuronal organisation is not ``strictly layered'' as for the network in Fig.~\ref{deepnet}, in which layer n neurons' inputs are only connected with layer n-1 neurons' outputs. In this case layer n neurons' inputs are connected with the outputs of neurons of all previous layers, down to sensory input neurons. We assume that the ``cabling'' linking all cortical neurons is laid out during embryonic development, before learning starts.  

\ifhpar \colorbox{colhd}{cortical neurons, radial connections} \\ \fi
The axons of all cortical neurons project also to the hippocampus, where their tips are very close to each other, much closer than their respective cellular bodies (also these connections are pre-set before the start of the learning process). When a set of neurons fire together, their axons make synaptic connections to the dendrites of a hippocampal neuron generated in the dentate gyrus (Fig.~\ref{hippoc}): in this way, a new record of the ``brain extended dataset'', encoding the co-occurrence of cortical activations, is created. 

\ifhpar \colorbox{colhd}{subset allowed to write /addition} \\ \fi
At each moment in the course of life, only a subset of neurons can write in the hippocampal dataset: in the beginning, this is done by sensory neurons only as all other neurons are not optimised. As upper layers' neurons become optimised, they also start recording their activations in the hippocampus. In the case of visual learning, for example, the hippocampal dataset is initially written by sensory neurons representing pixel intensity values. Once the concepts of digits become available as a result of learning, they are employed to create future records.

\ifhpar \colorbox{colhd}{subset allowed to write /deletion} \\ \fi
We may also hypothesise that ``obsolete'' neurons are gradually replaced by newer neurons and their axons gradually lose access to entorhinal cortex and hippocampus. This is consistent with the limited ``bandwidth'' allowed by the entorhinal cortex, whose size (presumably) only allows a subset of cortical axons to reach the hippocampus. In the case of visual learning, once the concepts of digits become available, neurons encoding individual retinal intensity values would cease to be used to create new associations (and possibly be physically disconnected). 

\ifhpar \colorbox{colhd}{need for convergence} \\ \fi
As already suggested in \citep{squire1986}, having all neurons project their axons to a small region is the only method to realise a fast association between features encoded in neurons that can be dispersed across a vast (in brain's terms) cortical area. If the neurons are distributed on the surface of a sphere (as it is in the brain's case), the best solution is that their axons project to a central point. In this point, the tips of all axons are very close to each other and the establishment of connections between them can be done very quickly. And it \textit{must} be done quickly, to keep up with the rapid flow of information fed from the perceptual apparatus. 

\ifhpar \colorbox{colhd}{role of hippocampus} \\ \fi
Based on these considerations, we suggest that the function of the hippocampus is to register and store the brain extended dataset, structured as an array of records composed of features, encoded in cortical neurons, that co-occurred. From this records, the ``algorithms'' of the cortex extract high-level features, encoded in other cortical neurons through adjustments of their lateral connections. Based on the two-phase algorithmic structure described in section 2, we assume that the optimisation procedure requires the continuous access to the hippocampal formation by cortical neurons, in both reading and writing. The generation of new features and the addition of new records to the hippocampus are continuous processes that run in parallel.

\ifhpar \colorbox{colhd}{instantiation of cortical memories} \\ \fi
Cortical memories are instantiated from the beginning in immature form, and gradually develop to more mature forms, exactly as described in \citep{kitamura2017}. For some episodic memories, that are essentially random associations of features without any hidden rules (the car can be parked near a tree or in front of a building without any deep reason other than the availability of a parking place), the process of ``cortical transfer'' can take longer, and essentially consists in copying to the cortex some salient elements of the otherwise untouched hippocampal record. 

\begin{figure}[t]
\begin{minipage}[t]{0.50\textwidth} \centering \hspace*{-0.0cm}
\includegraphics[width=1.00\textwidth]{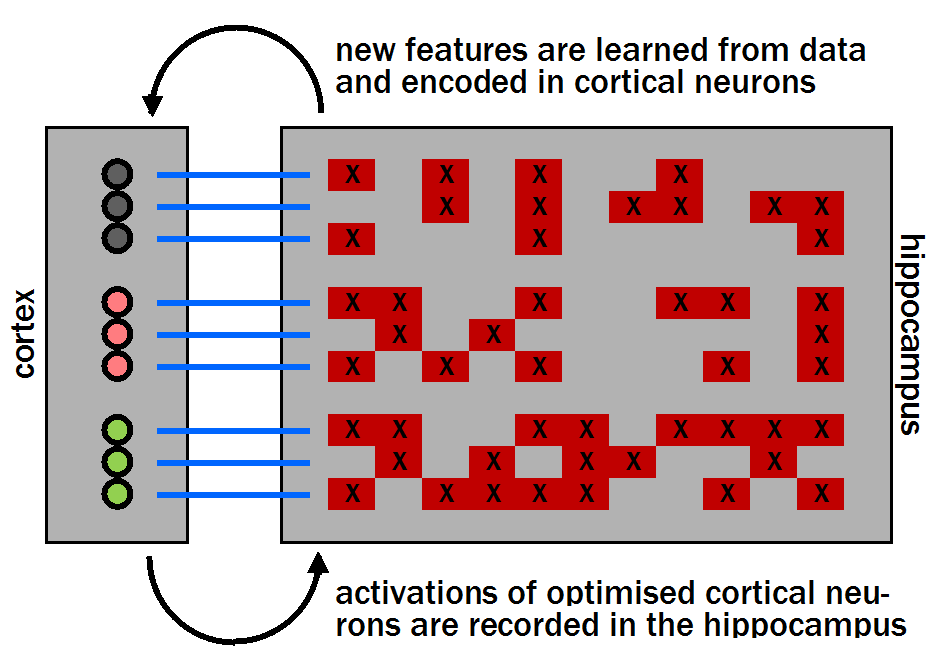}
\caption{Interplay between cortex and hippocampus. Cortical neurons register their activations in the hippocampus, whose records keep track of which neurons' activations co-occurred. In parallel, high-level features are learned and encoded in cortical neurons. In the beginning, sensory neurons write in the hippocampus while neurons of the first layers are optimised.}
\label{interplb}
\end{minipage} \qquad
\begin{minipage}[t]{0.50\textwidth} \centering \hspace*{-0.0cm}
\includegraphics[width=1.00\textwidth]{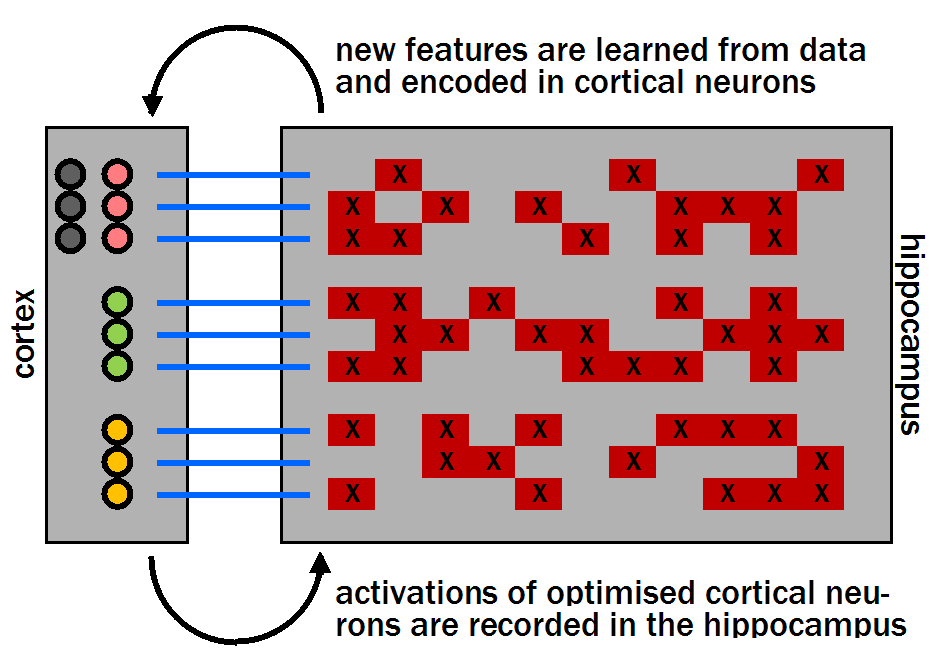}
\caption{Interplay between cortex and hippocampus. At a later stage, sensory neurons lose access to the hippocampus, replaced by upper layers' neurons, while neurons of even higher layers are optimised in the cortex.}
\label{interpll}
\end{minipage} \qquad
\end{figure}

Figs.~\ref{interplb} and ~\ref{interpll} show the interplay between cortex and hippocampus that results in learning and memory formation. Cortical neurons register their activations in the hippocampus, whose records keep track of which neurons' activations co-occurred. In parallel, high-level features are learned and encoded in cortical neurons. The subset of cortical neurons that have access to the hippocampus is continuously updated, adding new optimised neurons and removing obsolete neurons.   

  
\section{Discussion}  

\ifhpar \colorbox{colhd}{hypothesis consistent } \\ \fi
Our model of the hippocampus as the extended dataset of the brain is consistent with the finding the hippocampal neurons do more than encoding the spatial context \citep{eichenbaum2014} and goes beyond: we think that hippocampal neurons not only include space and time features, but all features. The model also accounts for the involvement of the hippocampus in flexible cognition and social behaviour \citep{rachael2014}, a task that requires the continuous updating of information and presupposes the continuous access to the hippocampus in both reading and writing.    

\ifhpar \colorbox{colhd}{connectionist models /1} \\ \fi
Hypotheses on the functioning of biological memory based on knowledge derived from connectionist models have already been proposed. \citet{mcclelland1995} carry out a very thorough analysis of the subject and conclude that ``interleaved learning'' is the most plausible model for the kind on learning that takes place in the brain. The authors argue that ``neural networks or connectionist models adhere to many aspects of the account of the mammalian memory system, but do not incorporate a special system for rapid acquisition of the contents of specific episodes and events''. 

\ifhpar \colorbox{colhd}{connectionist models /2} \\ \fi
Our counterargument to this observation is that neural networks do have a ``special system for rapid acquisition of the contents of specific episodes and events'' and this is nothing else than the computer memory section where the extended dataset is stored. The motivation for the failed recognition of this component is probably the notion that the data are strictly not part of the neural network model: they are however part of the system. 

\ifhpar \colorbox{colhd}{procedural memory /1} \\ \fi
H.M. had both hippocampal regions removed and, as a result, was unable to form new declarative memory: however, he was able to acquire new procedural memory. Different kinds of tasks fall under the definition of procedural memory: repetition priming, classical conditioning (Pavlov's experiments), emotional conditioning, various skills and habits such as mirror tracing, mirror reading or jigsaw puzzles. H.M.'s performance was good in many of these tasks \citep{woodruff1993}.

\ifhpar \colorbox{colhd}{procedural memory /2} \\ \fi
A first explanation for the fact that procedural learning can take place without hippocampus is that procedural tasks involve other brain regions (e.g., sensory cortex, cerebellum, amygdala, striatum). A second explanation is that the presence of a stored dataset is not always required. Learning may still be possible by processing each new record (or a small number or records, depending on the storage capacity available without hippocampus) ``on the fly'', changing the neurons' parameters and then testing the performance of the modified neurons on subsequent records. This would explain the ability of amnesiacs to learn often-repeated material gradually over time \citep{mcclelland1995, glisky1986}.

\ifhpar \colorbox{colhd}{hippocampus in PTSD} \\ \fi
The model proposed can have implications for the field of psychology: structural alterations to the hippocampus are in fact present in post-traumatic stress disorder \citep{bremner2006}, a condition caused by exposure to traumatic experiences that affects war veterans, victims of violence and abuse, etc. A common post-traumatic symptom is represented by dissociation, defined as the limitation or loss of the normal associative links between perceptions, thoughts and emotions. Dissociation can take the form of mental ``black-out'', depersonalisation, derealisation, selective amnesia and emotional detachment \citep{lanius2015, radovic2002}.

\ifhpar \colorbox{colhd}{interpretation: thin dataset} \\ \fi
Since the hippocampus is the device that provides the initial association among co-occurring features, it is natural to think that this structure is also involved in the absence of association. Dissociative symptoms can be modelled as a negative modulation of the links fed to the hippocampus (as well as to other brain regions), leading to a shrinkage of their target areas. Based on our model, such links are the carriers of as many cognitive features, and their disconnection translates to having an extended dataset ``thinner'' than normal (fewer rows, with reference to Fig.~\ref{dataset}). 

\ifhpar \colorbox{colhd}{hippocampus in schizophrenia} \\ \fi
Also the hippocampus of schizophrenic subjects is significantly reduced in size, and seems to be less susceptible to the phenomenon of habituation \citep{williams2013}: these observations are consistent with the hypothesis of psychosis as a ``learning and memory problem'' \citep{tamminga2013}. However, it is not clear whether such alterations are present also before the onset of symptoms or develop afterwards, as a result of disease progression. 

\ifhpar \colorbox{colhd}{unreliable associations in psychosis} \\ \fi
One of the most characteristic symptoms of psychosis is represented by delusions, which are beliefs held with strong conviction despite evidence to the contrary. For instance, if a person with psychosis sees one day two red cars parked near home, he/she may think that these cars are part of a conspiracy plot organised by a foreign government to spy and secure important industrial secrets: a conclusion that sounds absurd to most (non-psychotic) people. However, what would a normal person think if he/she saw two red cars parked next to the person's car each day for one month? Also this person would probably develop the idea of being followed.

\ifhpar \colorbox{colhd}{unreliable associations due to short dataset} \\ \fi
Therefore, the problem seems to be that the psychotic person jumps to the conclusion after an insufficient number of observations: as a result, the conclusion lacks statistical robustness (and is wrong). This could be caused by a smaller hippocampus able to host fewer data records. Based on our model, this translates to having an extended dataset ``shorter'' than normal (fewer columns, with reference to Fig.~\ref{dataset}): as a result, the cortical algorithms would be forced to learn from fewer records, leading to the establishment of unreliable associations. This, together with the phenomenon of aberrant salience \citep{howes2016}, could help to explain the nature of delusions in schizophrenia.  

\section{Conclusions}

\ifhpar \colorbox{colhd}{xxxx} \\ \fi
The objective of this work was to use the knowledge derived from the field of neural networks to gain insight into the functioning of biological memory and learning. Our central hypothesis is that the hippocampus is the biological device to store the brain extended dataset. Following this idea, memory and learning in neural networks appear fully compatible with the structure of the brain and the latest discoveries on biological memory. The model proposed can help to explain some psychological conditions, such as post-traumatic stress disorder and schizophrenia, in which the hippocampus presents structural alterations. 

\section{Disclaimer}

The author is an employee of the European Research Council Executive Agency. The views expressed are purely those of the writer and may not in any circumstances be regarded as stating an official position of the European Commission.

\bibliographystyle{apalike}
\bibliography{lhippostor}

\end{document}